\documentstyle[preprint,aps]{revtex}

\draft
\begin{document}
%\twocolumn[\hsize\textwidth\linewidth\hsize\csname @twocolumnfalse\endcsname

\title{Generalized master equation with nonhermitian operators}

\author{C. F. Huang$^{a,b}$ and K.-N. Huang$^{a,c}$}

\address{(a) Department of Physics, National Taiwan University, Taipei,
Taiwan, Republic of China \newline
(b)National Measurement Laboratory, Center for Measurement Standards, Industrial
Technology Research Institute, Hsinchu, Taiwan, Republic of China \newline
(c) Institute of Atomic and Molecular Sciences, 
Academia Sinica, P. O. Box 23-166, Taipei, Taiwan, Republic of China}

\date{\today}

\maketitle

\begin{abstract}

By extending the mean-field Hamiltonian to include nonhermitian operators,
the master equations for fermions and bosons can be derived. The derived
equations reduce to the Markoff master equation in the low-density limit
and to the quasiclassical master equation for homogeneuos systems.


\end{abstract}

\newpage

%\vskip2pc]

%\begin{multicols}{2}

\section{Introduction}

An adequate description for low-energy phenomena of fermions is provided by the
mean-field approximation in which particles move under the mean-field
Hamiltonian; relaxation effects are not taken into account in this
approximation [1]. The relaxation effects due to the reservoir, which
are important to the irreversibility [2], have been discussed in the
literature [1-7]. It was shown that loss and gain should be considered for such
effects and that the loss comes from the imaginary part of the self energy[4,8] in
the quantum master equation[6].

To include the self energy, it is reasonable to extend the mean-field
Hamiltonian to include nonhermitian operators. It will be shown in section II
that the master equation for fermions can be derived from such an extension, if
we assume that the equation is symmetric with respect to particles and
holes. The corresponding equation for bosons is obtained in section III. The
nonlinear master equation is obtained in section IV. To compare with other
stochastic equations, it is shown in section V that the derived equations
reduce to the Markoff master equation [3-5,7] in the low-density limit, and to the
quasiclassical master equation[9-11] for homogeneous systems. Conclusions are
made in section VI.

\section{Nonhermitian operators for fermions}

In the mean-field approximation, the system of identical particles is
described by a one-particle density matrix $\rho _{p}(t)$ whose trace equals the number of 
particles.[1] To extend such an approximation, we consider the quantum relaxation effects. 
For fermions, it will be shown in this section that the equation, including the relaxation 
term in addition to the Liouville flow, should be of the general form,
\begin{eqnarray}
\frac{\partial }{\partial t}\rho _{p}(t)=\frac{1}{i}[H(t),\rho _{p}(t)]+\{\rho
_{p}(t),A_{p}(t)\}-\{I-\rho _{p}(t),A_{\overline{p}}(t)\},
\end{eqnarray}
where $A_{p}(t)$ and $A_{\overline{p}}(t)$ are hermitian operators arising from
the relaxation effect, $H(t)$ is the hermitian operator for the Liouville flow,
and $I$ is the identity operator. Here we take $\hbar =1$, and notations $[A,B]$
and $\{A,B\}$ denote, respectively, the commutator and anticommutator of operators 
$A$ and $B$. As shown in section IV, from Eq. (1) we can intuitively obtain the relaxation 
term, which is formally derived in Appendix A from the second quantization.  

In the mean-field approximation, $\rho _{p}(t)$ is governed by
\begin{eqnarray}
\rho _{p}(t)=U(t,t^{\prime })\rho _{p}(t^{\prime })U^{\dagger
}(t,t^{\prime }),
\end{eqnarray}
with the time-evolution operator $U(t,t^{\prime })$ generated by the mean-field
Hamiltonian $H(t)$, which is a hermitian operator. To include the self energy
with imaginary part, it seems reasonable to extend $H(t)$ to a
Hamiltonian ${\cal H}(t)$ with a nonhermitian part,
\begin{eqnarray}
{\cal H}(t)=H(t)+iA(t)
\end{eqnarray}
where $A(t)$ is a hermitian operator. Then Eq. (2) is equivalent to
\begin{eqnarray}
\frac{\partial }{\partial t}\rho _{p}(t)=\frac{1}{i}[H(t), \rho _{p}(t) ]+\{\rho
_{p}(t),A(t)\}.
\end{eqnarray}
The first two terms on the right-hand side of Eq. (1) have been included in
Eq. (4) when we set $A(t)=A_{p}(t)$. However, if we assume the equation for
fermions to be symmetric with respect to particles and holes, Eq. (4) must
be modified to become Eq. (1), as shown later. The last two terms in Eq. (1) are
for the loss and gain factors of particles.

Instead of ${\rho }_{p}(t)$, fermions can be described as well by holes, i.e.,
the vacencies of particle orbitals [10]. First, we diagonalize
${\rho }_{p}(t)$ in an orthonormal basis $\{|\psi_{\lambda} (t)\rangle \}$:
\begin{eqnarray}
\rho _{p}(t)=\sum_{\lambda }c_{\lambda }(t)|\psi _{\lambda }(t)\rangle
\langle \psi _{\lambda }(t)|,
\end{eqnarray}
where $c_{\lambda}(t)$ denotes the number of particles in $|\psi _{\lambda
}(t)\rangle $. Define
\begin{eqnarray}
\rho _{\overline{p}}(t)\equiv I-\rho _{p}(t)=\sum_{\lambda }(1-c_{\lambda
}(t))|\psi _{\lambda }(t)\rangle \langle \psi _{\lambda }(t)|.
\end{eqnarray}
For any normalized state $|\phi \rangle $,
\begin{eqnarray}
\langle \phi |\rho _{\overline{p}}(t)|\phi \rangle =1-\langle \phi |\rho
_{p}(t)|\phi \rangle,
\end{eqnarray}
which is just the number of holes in that state, and therefore it is reasonable
to use $\rho _{\overline{p}}(t)$ to represent holes. In fact, we can obtain the
density matrix of holes from $\rho _{\overline{p}}(t)$, as shown in Appendix B.
In terms of  $\rho _{\overline{p}}(t)$, Eq. (4) becomes
\begin{eqnarray}
\frac{\partial }{\partial t}\rho _{\overline{p}}(t)=\frac{1}{i} [H(t), \rho 
_{\overline{p} }(t)]-\{I-\rho _{\overline{p}}(t),A(t)\}.
\end{eqnarray}
However, the form of the equation for $\rho_{\overline{p}}
(t)$ is different from that for $\rho_{p} (t)$ if $A(t)\not= 0$.

It is known that in Eqs. (4) and (8), $H(t)$ describes the Liouville flow.
To see the physical meaning of $A(t)$ in these two equations, let us
consider the special case that $H(t)=0$ and $A(t)=-\frac{\gamma (t)}{2}|\phi
\rangle \langle \phi |$ with $|\phi \rangle $ being a normalized state. Here
$\gamma (t)$ is real because $A(t)$ is hermitian.
From Eqs. (4) and (8) we obtain for particles and holes, respectively,
\begin{eqnarray}
\frac{\partial }{\partial t}\langle \phi |\rho _{p}(t)|\phi \rangle =-\gamma
(t)\langle \phi |\rho _{p}(t)|\phi \rangle,
\end{eqnarray}
\begin{eqnarray}
\frac{\partial }{\partial t}\langle \phi |\rho _{\overline{p}}(t)|\phi
\rangle =\gamma (t)(1-\langle \phi |\rho _{\overline{p}}(t)|\phi \rangle ),
\end{eqnarray}
When $\gamma(t)>0$, Eq. (9) describes the loss of particles in $| \phi \rangle$.
The loss of particles is just the gain of holes, and therefore Eq. (10) should
describe the gain of holes in $| \phi \rangle$ if $\gamma (t) >0$.

It seems natural to set $\gamma (t)<0$ in Eq. (9) to describe the gain of
particles in $|\phi \rangle$. The resulting equation, however, is not of the same
form as Eq. (10) with $\gamma (t)>0$ for the gain of holes. In fact, it seems natural to use
\begin{eqnarray}
\frac{\partial }{\partial t}\langle \phi |\rho _{p}(t)|\phi \rangle =\gamma
^{\prime }(t)(1-\langle \phi |\rho _{p}(t)|\phi \rangle )
\end{eqnarray}
to describe the gain of particles, just as we describe the gain of holes using
Eq. (10), where $ \gamma ^{\prime} (t)$ is a positive real number. Similarly, if
the loss of holes is described by Eq. (10) with $\gamma (t) <0$, the loss of
particles should also be described by Eq. (11) with $\gamma^{\prime} (t)<0$.

To correctly describe loss and gain of particles, we note that the loss rate of
particles should be zero when $|\phi \rangle $ has no particle, while from Pauli effect, in  
fact, the gain rate of particles should be zero when the orbital
is filled with particles. Therefore, Eq. (9) is only for the loss of
particles with $\gamma (t)>0$ while the gain of particles is described by
Eq. (11) with $\gamma ^{\prime }(t)>0$. Now the loss and gain rates are
proportional to $\langle \phi |\rho _{p}(t)|\phi \rangle $ and $1-\langle
\phi |\rho _{p}(t)|\phi \rangle $, respectively. In Appendix C it is shown
that Eq. (4) cannot describe the gain of particles in $|\phi \rangle $ no
matter what $H(t)$ and $A(t)$ are.

Because Eqs. (9) and (10) are obtained from the second terms on the right-hand
side of Eqs. (4) and (8), respectively, we can see why $\rho _{p}(t)$ should be
governed by Eq. (1), in which $A_{p}(t)$ and $A_{\overline{p}}(t)$ are for the
loss and gain of particles. Eq. (11) may be obtained from Eq. (1)
with $H(t)=A_{p}(t)=0$ and $A_{\overline{p}}(t)=-\frac{ \gamma^{\prime} (t) }{2}
|\phi \rangle \langle \phi |$, while Eq. (9) is obtained with $H(t)=A_{
\overline{p}}(t)=0$ and $A_{p}(t)=-\frac{\gamma (t)}{2}|\phi \rangle \langle
\phi |$. We can see that $A_{ \overline{p}}(t)$ is for the loss of holes just as
$A_{p} (t)$ for the loss of particles, by rewriting Eq. (11) with corresponding
$\rho _{\overline{p} }(t)$,
\begin{eqnarray}
\frac{\partial }{\partial t} \langle \phi | \rho _{\overline{p}}(t) | \phi \rangle =- 
\gamma ^{\prime} (t) \langle \phi |\rho _{\overline{p}}(t)|\phi \rangle .
\end{eqnarray}

Rewriting Eq. (1) with corresponding $\rho _{\overline{p}}(t)$, we arrive at
\begin{eqnarray}
\frac{\partial }{\partial t}\rho _{\overline{p}}(t)=  \frac{1}{i}[H(t), \rho 
_{\overline{p}}(t)]+\{\rho _{\overline{p}}(t),A_{\overline{p}}(t)\}-\{I-\rho _{%
\overline{p}}(t),A_{p}(t)\},
\end{eqnarray}
which is symmetric to Eq. (1) for $\rho_{p} (t)$.
To include an antihertmitian part in addition to the mean-field Hamiltonian
$H(t)$, therefore, we need to include $A_{\overline{p}}(t)$ in addition to $A_{p} (t)$. 
In section IV, the quantum master equation for fermions will be derived
from Eq. (1). Although all $c_{\lambda }(t)$ in Eq. (5) are 0 or 1 in the
mean-field approximation, they could be fractions in the master equation [1,2].

\section{The relaxation term for bosons}

For a boson system described by the one-particle density matrix $\rho_{p} (t)$,
it is reasonable that the decay rate of particles is proportional to $\langle
\phi | \rho _{p} (t) | \phi \rangle $ for any normalized state $| \phi \rangle$.
Unlike fermions, however, the gain rate for bosons may not vanish
when $\langle \phi | \rho _{p} (t) | \phi \rangle =1$, and hence we need
to modify the equation for the gain of particles. By using the equation
\begin{eqnarray}
\frac{\partial }{\partial t}\rho _{p}(t)=-\{ I + \rho
_{p}(t) , A _{\overline{p}} (t) \},
\end{eqnarray}
we have for the gain of particles in any arbitrary normalized ket
$| \phi \rangle$
\begin{eqnarray}
\frac{\partial}{\partial t} \langle \phi | \rho_{p} (t) | \phi \rangle =
\gamma(t) (1 + \langle \phi | \rho_{p} (t) | \phi \rangle),
\end{eqnarray}
where we set $A_{\overline{p}} (t) =  - \frac{\gamma (t)}{2} | \phi \rangle
\langle \phi |$ with a real number $\gamma (t)>0$. Now the rate for bosons to
enter $| \phi \rangle$ is proportional to $ 1 + \langle \phi | \rho _{p} (t) |
\phi \rangle $. The enhancement of the gain rate with increasing $\langle \phi |
\rho _{p} (t) | \phi \rangle$ is consistent with the fact that bosons
prefer states filled with many particles.

Therefore, for bosons we shall consider the following equation
\begin{eqnarray}
\frac{\partial }{\partial t}\rho _{p}(t)=\frac{1}{i}[H(t), \rho _{p}(t)]+\{\rho
_{p}(t),A_{p}(t)\}-\{I+\rho _{p}(t),A_{\overline{p}}(t)\}
\end{eqnarray}
to include the effects due to relaxation if Eq. (14) is suitable for the gain of bosons. 
As shown in the next section, from Eq. (16) we can intuitively obtain the 
relaxation term which is formally derived in Appendix A from the second quantization. Hence 
the modification for the gain of bosons is reasonable.

\section{Nonlinear quantum master equation}

In the last two sections, we obtain Eqs. (1) and (16) so that the loss and gain
factors are included in addition to the Liouville flow. To obtain the master
equation, first let us consider a two-state system in contact with the reservoir
and assume a time-independent Hamiltonian $H(t)=H_{0}$. Let $| i \rangle$
and $|f\rangle$ be normalized eigenkets of $H_{0}$. If the reservoir induces a
transition so that particles in $|i \rangle$ tend to jump to $|f\rangle$,
there should be a loss factor for $|i\rangle$ and a gain factor for $|f\rangle$.
Therefore for $A_{\overline{p}} (t)$ and $A_{p} (t)$ in Eqs. (1) and (16),
respectively, we shall set
\begin{eqnarray}
A_{\overline{p}}^{(fi)}(t)=-\frac{{\gamma }_{1}(t)}{2}|f\rangle \langle f|,
\end{eqnarray}
\begin{eqnarray}
A_{p}^{(fi)}(t)=-\frac{{\gamma }_{2}(t)}{2}|i\rangle \langle i|,
\end{eqnarray}
where $\gamma_{1} (t)$ and $\gamma_{2}(t)$ are positive real numbers. So we have
\begin{eqnarray}
\frac{\partial }{\partial t}\langle f|\rho _{p}(t)|f\rangle ={\gamma}_{1} (t)
( 1 \pm \langle f|\rho _{p}(t)|f\rangle),
\end{eqnarray}
\begin{eqnarray}
\frac{\partial }{\partial t}\langle i|\rho _{p}(t)| i \rangle =-{\gamma }_{2}
(t) \langle i | \rho _{p}(t) | i \rangle .
\end{eqnarray}
In the right-hand side of Eq. (19), the $\lq \lq + "$ and $\lq \lq - "$ signs
are for bosons and fermions, respectively. From the conservation of the number
of particles, we have
\begin{eqnarray}
{\gamma }_{1}(t)( 1 \pm \langle f|\rho_{p}(t)| f \rangle )= {\gamma}_{2} (t)
\langle i|\rho _{p} (t) | i \rangle .
\end{eqnarray}
We can set a nonnegative number $w_{fi}(t)$ so that $\gamma_{1} (t) = w_{fi} (t)
\langle i|\rho _{p}(t)|i\rangle$ and $\gamma_{2} (t) = w_{fi} (t) (1 \pm \langle
f|\rho _{p}(t)|f\rangle )$ in order to satisfy Eq. (21). Then we have
\begin{eqnarray}
A_{ \overline{p} }^{(fi)}(t)= -\frac{1}{2} w_{fi} (t) \langle i | \rho_{p} (t) |
i \rangle | f \rangle \langle f|,
\end{eqnarray}
\begin{eqnarray}
A_{p}^{(fi)} (t) =-\frac{1}{2} w_{fi}(t) (1 \pm \langle f|\rho _{p}(t) | f \rangle)
|i\rangle \langle i|.
\end{eqnarray}
Taking $A_{ \overline{p} }^{(fi)}(t)$ and $A_{p}^{(fi)}(t)$ as $A_{p} (t)$ and 
$A_{\overline{p}}(t)$, we can derive the master equation for such a special case after 
inserting them into Eqs. (1) and (16). Now Eqs. (19) and (20) become  
\[
\frac{\partial}{\partial t} \langle f | \rho _{p} (t) | f \rangle = - \frac{\partial}{
\partial t} \langle i | \rho _{p} (t) | i \rangle = w _{fi} (t) \langle i | \rho _{p} (t) 
| i\rangle (1 \pm \langle f | \rho _{p} (t) | f \rangle ).
\]    
In the above equation, the change rates for bosons and fermions are different because of 
the last factor, which is determined by the number of particles in $|f \rangle$.
For bosons the transition becomes faster if there are more particles in 
$| f \rangle$. On the other hand, for fermions the transition is forbidden 
when $| f \rangle$ is filled with particles, which is consistent with Pauli effects.

In general, the reservoir may induce many transitions when the system contains
many orbitals. Assume that the transitions are within an orthonormal complete
set $S$ in which each orbital $n$ corresponds to a normalized ket $|n\rangle$.
The transition $n \rightarrow n^{\prime}$ should induce the loss and gain in $n$
and $n^{\prime}$ without changing the number of particles; we shall set
$A_{\overline{p}}^{(n^{\prime}n)}(t)=-\frac{1}{2} w_{n^{\prime}n} (t) \langle
n|\rho_{p} (t) | n \rangle | n^{\prime} \rangle \langle n ^{\prime} |$ and $
A_{p}^{(n^{\prime} n)}(t) = - \frac{1}{2} w _{n^{\prime} n} (t) ( 1 \pm \langle
n^{\prime} | \rho _{p} (t) | n^{\prime} \rangle ) | n \rangle \langle n | $ for
such a transition. To include all transitions, we shall set $ A_{\overline{p}}
(t) = \sum_{(n^{\prime}n)}A_{\overline{p}}^{(n^{\prime}n)}(t)$ and $A _{p} (t)=
\sum_{(n^{\prime}n)}A_{p}^{(n^{\prime}n)}(t)$: Eqs. (1) and (16) become
\begin{eqnarray}
\frac{\partial}{\partial t} \rho_{p} (t) = \frac{1}{i}[ H(t), \rho_{p} (t)]+
\sum_{(n^{\prime}n)}\{ \rho_{p}(t),A_{p}^{( n^{\prime}n)}(t)\} - \sum_{( n^{
\prime}n) } \{ I \pm \rho_{p} (t),A_{ \overline{p} }^{ (n^{\prime} n) } (t) \},
\end{eqnarray}
\[
=\frac{1}{i}[H(t), \rho_{p} (t)]- \frac{1}{2}\sum_{(n^{\prime}n)} w_{n^{\prime}n} (t) (1
\pm \langle n^{\prime} | \rho _{p} (t) | n ^{\prime} \rangle ) \{ \rho _{p} (t),
| n \rangle \langle n | \}
\]
\[
+\frac{1}{2}\sum_{(n^{\prime}n) } w_{n^{\prime}n} (t) \langle n | \rho _{p} (t)|
n \rangle \{ I \pm \rho _{p} (t), | n ^{\prime} \rangle \langle n ^{\prime} |\},
\]
which is a nonlinear equation. Comparing the above equation to Eq. (A16) derived in 
Appendix A, we can see that the relaxation term can also be obtained formally from the 
second quantization, by considering a system composed of noninteracting particles in 
contact with the reservoir. 

\section{Discussions}

For stochastic processes, several equations are used under different
conditions [1-11]. It will be shown in this section that we can derive the
Markoff master equation [3-5,7] and quasiclassical master equation [9-11]
from Eq. (24). An extension of Eq. (24) is also obtained after comparing Eq.
(24) with the quantum master equation [6].

Consider the homogeneous case first. That is, the Hamiltonian and the density
matrix commute with the momentum operators ${\bf P}$, and $S$ is composed of
the eigenlevels of ${\bf P}$. Let $|{\bf p}\rangle$ be the plane wave with the
momentum ${\bf p}$. By setting $f({\bf p},t)\equiv \langle {\bf p} |
\rho_{p} (t)|{\bf p} \rangle$, we obtain from Eq. (24)
\begin{eqnarray}
\frac{\partial }{\partial t}f({\bf p},t)=\sum_{{\bf p}^{\prime }}w_{{\bf p}{\bf
p}^{\prime }}(t) [ 1\pm f({\bf p},t) ] f({\bf p}^{\prime},t)
\end{eqnarray}
\[
- \sum_{{\bf p}^{\prime }}w_{{\bf p}^{\prime } {\bf p}} (t) [ 1 \pm f( {\bf
p}^{\prime },t) ] f({\bf p},t),
\]
which is just the quasiclassical master equation.[2,10]

Next we consider the low-density limit. If we diagonalize $\rho _{p}(t)$ as
\[
\rho_{p} (t) = \sum_{\lambda} c_{\lambda} | \psi_{\lambda} (t) \rangle \langle
 \psi_{\lambda} (t) |,
\]
then all $c_{\lambda }(t)$ are small, such that $I \pm \rho _{p} (t) =
\sum_{\lambda } [ 1 \pm c_{\lambda} (t) ] |\psi _{\lambda} (t) \rangle \langle
\psi _{\lambda } (t) | \simeq I$. In addition, $A _{p}^{(n^{\prime}n)} (t)$ in
Eq. (24) can be reduced to $- \frac{1}{2} \sum _{(n^{\prime} n)} w_{ n^{\prime}
n} (t) | n \rangle \langle n |$. If coeffecients $w_{n^{\prime} n} (t)$ are
independent of $t$, Eq. (24) is reduced to
\begin{eqnarray}
\frac{\partial }{\partial t}\rho_{p}(t)=\frac{1}{i}[H(t), \rho_{p}(t)]- \frac{1}{2}
\sum_{(n^{\prime}n)} w_{n^{\prime}n} \{ \rho_{p}(t), |n \rangle \langle n| \}
\end{eqnarray}
\[
+\sum_{ ( n^{\prime} n ) } w_{n^{\prime} n} \langle n | \rho _{p} (t) | n
\rangle |n ^{\prime} \rangle \langle n^{\prime} |.
\]
The above equation is just the so-called Markoff master equation [3-5,7]:
\begin{eqnarray}
\frac{\partial }{\partial t}\rho _{p}(t)=\frac{1}{i}[H(t),\rho_{p} (t)] - \frac{1}{2}
\sum_{( n^{\prime} n)} w_{n^{\prime}n} \{ \rho_{p} (t), |n \rangle \langle n| \}
\end{eqnarray}
\[
+\sum_{n^{\prime} n }w_{n^{\prime} n} \langle n|\rho _{p} (t) | n \rangle |
n^{\prime} \rangle \langle n^{\prime}|
-\sum_{ n \not= n^{\prime} } \Gamma_{ n^{\prime} n} \langle n | \rho _{p} (t) |
n^{\prime} \rangle | n \rangle \langle n^{\prime}|,
\]
with all $\Gamma _{n^{\prime} n} =0$. The last term of Eq. (27) is the
so-called pure-dephasing term. In the Markoff master equation, $\{ \Gamma_{
n^{\prime} n} \}$ is a set of nonnegative numbers with $\Gamma _{n^{\prime}n} =
{\Gamma }_{n n^{\prime}}$. It is easy to see that for fermions the
pure-dephasing term is symmetric with respect to particles and holes, and hence
it seems that we can add such term into Eq. (1). The positivity of $\rho_{p} (t)
$, however, can be broken under a pure-dephasing term, as shown in Appendix D.

Eq. (26) is, in fact, a particular case of the quantum master equation [6,12],
\begin{eqnarray}
\frac{\partial}{\partial t} \rho _{p}(t) = \frac{1}{i}[H(t), \rho _{p} (t)] - \frac{1}{2}
\sum_{l}\{\rho _{p}(t),{\cal W}_{l}{\cal W}_{l}^{\dagger } \}+\sum_{l}{\cal W
}_{l}^{\dagger }\rho _{p}(t){\cal W}_{l}
\end{eqnarray}
if we set
\begin{eqnarray}
{\cal W}_{l}=\sqrt{ w_{ n^{\prime} n} } |n\rangle \langle n^{\prime}|,
\end{eqnarray}
where $\{ w_{n^{\prime} n} \}$ is a set of positive constants and $l \equiv
(n^{\prime},n)$. Therefore, Eq. (28) can be taken as a generalization of Eq.
(26). Similarly, we can generalize Eq. (24) to
\begin{eqnarray}
\frac{\partial }{\partial t}\rho _{p}(t)=\frac{1}{i}[H(t),\rho _{p}(t)]-\frac{1}{2}
\sum_{l}\{\rho _{p}(t),{\cal W}_{l} (t) (I\pm \rho _{p}(t)){\cal W}_{l}^{\dagger
} (t) \}
\end{eqnarray}
\[
+ \frac{1}{2}\sum_{l}\{I\pm \rho _{p}(t),{\cal W}_{l}^{\dagger } (t) \rho
_{p}(t) {\cal W}_{l} (t) \}
\]
with an arbitrary set of operators $\{{\cal W}_{l} (t) \}$. Eq. (30) can be
obtained from Eqs. (1) and (16) by setting $A_{p}(t)=-\frac{1}{2}\sum_{l}
{\cal W}_{l} (t)(I \pm \rho_{p}(t)) {\cal W}_{l}^{\dagger} (t)$ and $A_{\overline{p}}(t)
=-\frac{1}{2} \sum_{l}{\cal W} _{l}^{\dagger } (t) \rho _{p}(t) {\cal W}_{l} (t)$, and 
it can be reduced to Eq. (28) in the low-density limit (under which $I \pm \rho _{p} (t) 
\simeq I$). Therefore, Eqs. (1) and (16) are general equations for quantum stochastic
processes. It is shown in Ref. 13 that under suitable assumptions the positivity
of $\rho_{p} (t)$ is preserved under Eqs. (24) and (30). In addition, for
fermions we have $ \langle \alpha | \rho _{p} (t) | \alpha \rangle \leq 1$ for
all normalized ket $| \alpha \rangle $, if such a relation holds initially.

As mentioned in section II, we shall include the second term of Eq. (1) if the 
time-evolution operator $U(t,t^{\prime})$ in Eq. (2) is generated by a nonhermitian Hamiltonian 
${\cal H}(t)$. For fermions we can operate such a time-evolution operator on the matrix 
$\rho_{\overline{p}}(t)$, which is defined in Eq. (6) for holes, and obtain the following 
equation    
\begin{eqnarray}
\frac{\partial }{\partial t}\rho _{\overline{p}}(t)=\frac{1}{i}[H(t), \rho _{\overline{p}%
}(t)]+\{\rho _{\overline{p}}(t), A(t) \}.
\end{eqnarray}
Replacing $A (t)$ by $A _{ \overline{p} } (t)$, we can see that the last term of Eq. (1) 
corresponds to the last term of the above equation, after rewriting Eq. (1) as Eq. (13).  
Therefore, by considering the nonhermitian Hamiltonian, Eq. (1) can also be derived by 
including the last terms in both Eqs. (4) and (31), while in section II it is derived from 
Pauli effects. Setting $A ^{\prime} (t) = A _{p} (t) \mp A _{ \overline{p} } (t)$ with 
the signs $ \lq \lq -"$ and $\lq \lq + "$ for bosons and fermions, actually 
we can rewrite Eqs. (1) and (16) as 
\begin{eqnarray}
\frac{\partial}{\partial t} \rho _{p} (t) = \frac{1}{i} [ H(t) , \rho _{p} (t)]
+ \{ \rho _{p} (t) , A ^{\prime} (t) \} - 2 A _{ \overline{p} } (t).
\end{eqnarray}
But it is easy to see that for fermions, Eq. (1) is of a symmetric form with respect to 
particles and holes, after rewriting Eq. (1) as Eq. (13). To compare the equations for bosons
and fermions, we shall use Eq. (16) for bosons, rather than the above one, when Eq. (1) is 
used for fermions.    

In the quasiparticle theory [14-16], which is a good approximation for 
many-electron systems, with suitable assumptions the wavefunction $\Psi (t)$ 
of a quasiparticle (or a quasihole) is governed by [16] 
\begin{eqnarray}
i \frac{\partial}{\partial t} | \Psi (t) \rangle = {\cal H} (t) | \Psi (t) \rangle. 
\end{eqnarray}
Here the nonhermitian Hamiltonian ${\cal H} (t)$ is defined in Eq. (3), and its nonhermitian 
part is due to the finite lifetimes of the quasiparticles (quasiholes). 
The above equation can correspond to Eq. (4) or Eq. (31) by rewriting it as
\begin{eqnarray} 
\frac{\partial}{\partial t} \rho ^{\prime} (t) = \frac{1}{i} [H(t), \rho ^{\prime} (t)]
+ \{ \rho ^{\prime} (t) , A (t) \}, 
\end{eqnarray}
with $\rho ^{\prime} (t) \equiv | \Psi (t) \rangle \langle \Psi (t) |$. 
From the quasiparticle theory, therefore, we can see why the nonhermitian part of the 
Hamiltonian should correspond to the decay of particles or that of holes. Since in an 
orbital the decay of particles (holes) is equivalent to the increase of holes (particles), 
it is natural to consider both the last terms of Eqs. (4) and (31) to obtain Eq. (1), 
as mentioned above. But it should be emphasized that in this paper particles and holes are, 
respectively, the filled and empty parts of any orbitals, while in the quasiparticle theory 
they are the filled and empty parts above and below the Fermi energy. While Eq. (1) can be 
applied to a nonequilibrium Fermi system with no well-defined Fermi energy, in the 
quasiparticle theory the system is not far away from the equilibrium and we shall still 
consider Fermi energy. In addition, Eq. (1) describes the time evolution of a whole system, 
but in the above equation we are only interested in the energy and lifetime of a 
quasiparticle or quasihole, and do not consider the third term of Eq. (1) or Eq. (13). 
 
\section{Conclusion}

It is shown in this paper that the master equation for fermions is of a
symmetric form with respect to particles and holes, and hence can be derived by
extending the mean-field Hamiltonian to include an antihermitian part. The
equation for bosons is also obtained. The derived equations reduce  to the
Markoff master equation in the low density limit and to the quasiclassical
master equation for homogeneous systems.

\section*{Appendix A: Markoff master equation}

Consider a system ${\cal S}$ coupled to a reservoir $R$. The total Hamiltonian
$H=H_{{\cal S}}+H_{R}+{\cal V}$, where $H_{{\cal S}}$, $H_{R}$, and ${\cal V}$
are the operators for the system, the reservoir, and the interaction between
${\cal S}$ and $R$, repectively. In addition, $[H_{{\cal S}},H_{R}]=0$. If the
system is composed of noninteracting identical particles, we can write
\[
H_{{\cal S}}=\sum_{n}\varepsilon _{n}c_{n}^{\dagger }c_{n}, \ \ \ \ \ \ \ \ \ \ \
\ \ \ \ \ \ (A1) 
\]
with $\{c_{n},c_{n^{\prime}}^{\dagger}\}=\delta _{n,n^{\prime} }$ and $\{ c_{n},
c_{ n^{\prime} } \}=0 $ for femions and $[c_{n},c_{ n^{\prime} }^{\dagger
}]=\delta _{n,n^{\prime} }$ and $[c_{n},c_{ n^{\prime} }]=0$ for bosons.

Assume ${\cal V}=\sum_{n\not=n^{\prime }}f_{n,n^{\prime }}c_{n}^{\dagger
}c_{n^{\prime }}$, where $f_{n,n^{\prime }}$ contain the operators for $R$
and satisfy $f_{n,n^{\prime }}=f_{n^{\prime },n}^{\dagger }$. Under suitable
assumptions [3,4,7] the density matrix $\rho _{R}$ of the reservoir can be
expanded by the eigenlevels $ | r \rangle $ of $H_{R}$:
\[
\rho _{R}=\sum_{r}F(E_{r})|r\rangle \langle r|.\ \ \ \ \ \ \ \ \ \ \
\ \ \ \ \ \ (A2) 
\]
Here $E_{r}$ is the eigenvalue corresponding to $|r\rangle $ and $F(E_{r})$
is the thermal equilibrium distribution. Let $\rho _{{\cal S}} (t) $ be the
density matrix of the system and $\rho _{C} (t)$ be the total density matrix, then
\[
\rho _{{\cal S}} (t) = \sum _{r} \langle r | \rho _{C} (t) | r \rangle.\ \ \ \ \ \ \ \ \ \ \
\ \ \ \ \ \ (A3) 
\]
In addition, at $t=0$ we assume that
\[
\rho _{C}(0)=\rho _{{\cal S}}(0)\otimes \rho _{R} \ \ \ \ \ \ \ \ \ \ \ \ \ 
\ \ \ \ \ \ (A4) 
\]
with positive $\rho _{{\cal S}} (0)$ satisfying the normalization condition
$tr\rho _{{\cal S}} (0) =1$.

Choose ${\cal H}_{0}=H_{{\cal S}}+H_{R}$ as the unperturbed Hamiltonian, the
time-evolution operator, ${\cal U}^{(I)}(t,t^{\prime})$, in the interaction picture
satisfies
\[
\frac{\partial }{\partial t}{\cal U}^{(I)}(t,t^{\prime })={\cal V}%
^{(I)}(t) {\cal U} ^{(I)} ( t , t^{\prime} ), \ \ \ \ \ \ \ \ \ (A5) 
\]
where ${\cal V}^{(I)}(t)=\sum_{n \not= n^{\prime} }f_{n,n^{\prime
}}^{(I)}(t)e^{i(\varepsilon _{n}-\varepsilon _{n^{\prime }})t}c_{n}^{\dagger
}c_{n^{\prime }}$ with $f_{n,n^{\prime }}^{(I)}(t)=e^{iH_{R}t}f_{n,n^{\prime
}}e^{-iH_{R}t}$. At $t>0$, we have
\[
\rho _{C}^{(I)}(t)={\cal U}^{(I)}(t)\rho _{C}(0){\cal U}^{(I)\dagger }(t).\ \ \ \ \ \ \ \ \ \ \ \ \ 
\ \ \ \ \ \ (A6) 
\]
in the interaction picture. After expanding $\rho _{{\cal S}}^{(I)}(t)$ to the
second order with respect to ${\cal V}$ and assume that $ Tr \rho _{R}
f_{n,n^{\prime }}=0$ [3,4] we have
\[
\rho _{{\cal S}}^{(I)}(t)-\rho _{{\cal S}}(0)=it[\rho _{{\cal S}}(0),\Sigma
_{R}]-t\{\rho _{{\cal S}}(0),\Sigma _{I}\}+\sum_{r,r^{\prime }}F(E_{r
})\int_{0}^{t}\int_{0}^{t}\times \ \ \ \ \ \ (A7) 
\]
\[
\sum_{m \not= m^{\prime},n \not= n ^{\prime} } dt_{1} dt_{2} e ^{ i [ (
\varepsilon _{n}- \varepsilon _{ n^{\prime} }) t_{1} +( \varepsilon _{m} -
\varepsilon _{m^{\prime }})t_{2}]}\langle r|f_{n,n^{\prime
}}^{(I)}(t_{1})|r^{\prime }\rangle \langle r^{\prime }|f_{m,m^{\prime
}}^{(I)}(t_{2})|r\rangle c_{n}^{\dagger }c_{n^{\prime }}\rho
_{{\cal S}}(0)c_{m}^{\dagger }c_{m^{\prime }},
\]
where $\Sigma _{R}$ and $\Sigma _{I}$ are the real and imaginary parts of
\[
\Sigma \equiv \frac{i}{t}\sum_{r,r^{\prime }}F(E_{r
})\int_{0}^{t}dt_{1}\int_{0}^{t_{1}}dt_{2}\langle r|{\cal V}%
^{(I)}(t_{1})|r^{\prime }\rangle \langle r^{\prime }|{\cal V}%
^{(I)}(t_{2})|r\rangle . \ \ \ \ \ \ (A8)
\]
If $t$ is much longer than the correlation time of the reservoir, $\Sigma $
is just the self energy [4,8].

To calculate the third term on the right hand side of Eq. (A7), let $\langle
r|f_{n,n^{\prime }}|r^{\prime }\rangle =$\ $|\langle r|f_{n,n^{\prime
}}|r^{\prime }\rangle |e^{i\theta _{n,n\prime }(r,r\prime )}$. Since $%
f_{n,n^{\prime }}=f_{n^{\prime },n}^{\dagger }$, $\theta $$_{n,n^{\prime
}}(r,r^{\prime })=-\theta _{n^{\prime },n}(r^{\prime },r)$. Assume that $%
\theta _{n,n^{\prime }}(r,r^{\prime })$ is random with respect to $r^{\prime
}$\, while $| \langle r|f_{n,n^{\prime }}| r ^{\prime} \rangle |$ varies slowly
with respect to $r^{\prime }$. In addition, when $ ( n, n ^{\prime} ) \not=
( m^{\prime}, m ) $, $\phi _{ n, n^{\prime} , m , m^{\prime} } ( r, r^{\prime} )
\equiv \theta _{ n, n^{\prime} }( r , r^{\prime} ) - \theta _{ m,m^{\prime} }
( r, r^{\prime } ) $ is still random. That is, the change of $ \theta _{n,
n^{\prime} } (r,r^{\prime} ) $ due to the reservoir is random, and there is no
correlation between different transitions. Under these assumptions, only those
terms with $(n,n^{\prime })=(m^{\prime },m)$ in the summation do not vanish.

After some calculations, we can set $w_{n^{\prime} n} =\sum _{ r, r^{\prime} }
F( E_{r} ) | \langle r | f_{ n,n^{\prime} } | r ^{\prime} \rangle | ^{2} \delta
_{ E_{n} + E_{r}, E_{ n^{\prime} } + E_{ r^{\prime} } }$ as the transition rate
for particles to jump from $n$ to $n^{\prime}$ so that
\[
\Sigma _{I}=\frac{1}{2}\sum_{n \not= n ^{\prime}} w_{n^{\prime} n} c_{n}^{
\dagger } c_{ n^{\prime} } c_{ n^{\prime} }^{ \dagger } c_{n}. \ \ \ \ \ \ \ \ \ \ \ \ \ 
\ \ \ \ \ \ (A9) 
\]
Ignoring $\Sigma _{R}$ [3,4], Eq. (41) in the Schr\"{o}dinger picture becomes
\[
\frac{\partial }{\partial t}\rho _{{\cal S}}(t)|_{t=0} = i[\rho_{{\cal S}} (t),
H_{{\cal S}}]-\frac{1}{2}\sum_{ n \not= n^{\prime} }w_{n^{\prime} n} \{ c _{n}
^{\dagger} c_{ n^{\prime} } c_{n^{\prime}}^{\dagger} c_{n},\rho_{{\cal S}}(t) \}
\ \ \ \ \ \ (A10)
\]
\[
+\sum_{n \not=n^{\prime} }w_{n^{\prime} n} c_{ n^{\prime} }^{\dagger} c_{n}
\rho _{{\cal S}}(t)c_{n}^{\dagger }c_{ n^{\prime} }.
\]
Assuming that the above equation holes at any time [3,4] and setting ${\cal A}_{
n^{\prime} n} = \sqrt{ w_{n^{\prime} n} }c_{n^{\prime}}^{\dagger}c_{n}$, we
can reduce the above equation to be
\[
\frac{\partial }{\partial t}\rho _{{\cal S}} (t) = \frac{1}{i} [H_{{\cal S}}, \rho _{{\cal S}}(t)]-
\frac{1}{2} \sum_{ n \not= n^{\prime} } \{ {\cal A}_{n^{\prime} n} ^{\dagger}
{\cal A}_{n^{\prime} n},\rho _{{\cal S}} (t) \} + \sum_{ n \not= n^{\prime} } {\cal
A}_{n^{\prime} n} \rho _{{\cal S}} (t) {\cal A}_{n^{\prime} n} ^{\dagger},
\ \ \ \ \ \ (A11)
\]
which is just the quantum master equation [6] with the second and third terms
as the loss and gain factors.

Let $| n \rangle $ be the orbital corresponding to $c_{n}$, and $\rho _{p} (t)$
be the one-particle density matrix satisfying $\langle n| \rho _{p} (t) |
n^{\prime} \rangle =Trc_{n}^{\dagger} c_{n^{\prime }} \rho _{{\cal S}} (t) $. From
Eq. (A10), the diagonal terms satisfy
\[
\frac{\partial }{\partial t}\langle n|\rho _{p}(t)|n\rangle =\sum_{n^{\prime
} \not= n}[-w_{n^{\prime} n}Tr(c_{n}^{\dagger }c_{n}c_{n^{\prime }}c_{n^{\prime
}}^{\dagger }\rho _{{\cal S}}(t))+w_{ n n^{\prime} }Tr(c_{n^{\prime }}^{\dagger
}c_{n^{\prime }}c_{n}c_{n}^{\dagger }\rho _{{\cal S}}(t))], \ \ \ \ \ \ (A12)
\]
and phases are governed by
\[
\frac{\partial }{\partial t}\langle n|\rho _{p}(t)|n^{\prime \prime }\rangle
=i(\varepsilon _{n^{\prime \prime }}-\varepsilon _{n})\langle n|\rho
_{p}|n^{\prime \prime }\rangle \ \ \ \ \ \ \ \ \ \ \ \ \ 
\ \ \ \ \ \ (A13) 
\]
\[
-\frac{1}{2}\sum_{n^{\prime} \not= n }w_{n^{\prime} n}Tr(c_{n^{\prime }} c_{n^{\prime }}
^{\dagger } c _{n} ^{\dagger} c _{n^{\prime \prime }} {\rho }_{{\cal S}} (t)) 
+\frac{1}{2} \sum_{ n^{\prime} \not= n } w_{n n^{\prime} } Tr(c_{n^{\prime }}^{\dagger }
c_{n^{\prime}}c_{n^{\prime \prime }} c_{n} ^{\dagger} \rho _{{\cal S}}(t)) 
\]
\[
+\frac{1}{2}\sum_{n^{\prime}\not= n^{\prime\prime}}w_{n ^{\prime \prime } n ^{
\prime} }Tr(c_{n^{\prime \prime }} c_{n} ^{\dagger} c_{n^{\prime }}^{\dagger
}c_{n^{\prime }} \rho _{{\cal S}}(t)) -\frac{1}{2} \sum_{ n^{\prime} \not= n ^{\prime 
\prime}} w_{n^{\prime} n ^{ \prime \prime } }
Tr(c_{n}^{\dagger } c_{n^{\prime \prime }} c_{n^{\prime}} c_{n^{\prime}} ^{\dagger}
\rho _{{\cal S}}(t)),
\]
where $n \not= n^{\prime \prime}$. 

Disjointing from Eq. (A12), so that $c_{n^{\prime }}$ and $c_{n^{\prime }}^{\dagger }$ are 
paired, we have 
\[
\frac{\partial}{\partial t} \langle n | \rho _{p} (t) | n \rangle
=-\frac{1}{2} \sum _{n ^{\prime} } w_{ n ^{\prime} n  }
(1 \pm \langle n ^{\prime} | \rho _{p} (t) | n ^{\prime} \rangle ) \langle n | \rho _{p} (t) 
| n \rangle  \ \ \ \ \ (A14)
\]
\[
+ \frac{1}{2} \sum _{ n ^{\prime} }  w _{ n n ^{\prime} }   ( 1 \pm \langle n | \rho _{p} 
(t) | n \rangle ) \langle n ^{\prime} | \rho _{p} (t) | n ^{\prime} \rangle,
\]
while from Eq. (A13) we have 
\[
\frac{\partial}{\partial t} \langle n | \rho _{p} (t) | n ^{\prime \prime} \rangle | _{n 
\not= n ^{\prime \prime}} =i \langle n |[ \rho _{p} (t),H_{0}]| n ^{\prime \prime } \rangle
\ \ \ \ \ \ \ \ \ \ \ \ \ \ \ \ \ \ \ (A15) 
\]
\[
-\frac{1}{2} \sum _{n ^{\prime} } (w_{ n ^{\prime} n ^{\prime \prime} } + w_{ n ^{\prime} n})
(1 \pm \langle n ^{\prime} | \rho _{p} (t) | n ^{\prime} \rangle ) \langle n | \rho _{p} (t) 
| n ^{\prime \prime} \rangle 
\]
\[
\pm \frac{1}{2} \sum _{ n ^{\prime} } ( w _{ n n ^{\prime} } + w _{ n ^{\prime \prime} n 
^{\prime} } )   \langle n | \rho _{p} (t) | n 
^{ \prime \prime } \rangle  \langle n ^{\prime} | \rho _{p} (t) | n ^{\prime} \rangle.
\]       
Here the signs $\lq \lq + "$ and $\lq \lq -"$ in the symbol $\lq \lq \pm "$ are 
for bosons and fermions, respectively. Although in Eq. (A13) the disjointness is not allowed 
for $n^{\prime}= n$ or $ n^{\prime \prime }=n$, it is suitable to ignore such a problem if 
the summation is over many terms. 

It has been shown that Eq. (A14) corresponds to a semiclassical master euqation. Although 
Eq. (A15) looks complicated, with some calculations we can see that $\rho _{p} (t)$ is 
governed by the following equation,     
\[
\frac{\partial}{\partial t} \rho_{p} (t) = \frac{1}{i}[ H _{0}, \rho_{p} (t) ]
- \frac{1}{2}\sum_{(n^{\prime}n)} w_{n^{\prime} n}  (1
\pm \langle n^{\prime} | \rho _{p} (t) | n ^{\prime} \rangle )\{ \rho _{p} (t),
| n \rangle \langle n | \}  \ \ \ \ \ (A16)
\]
\[
+\frac{1}{2}\sum_{(n^{\prime}n) } w_{n^{\prime}n} \langle n | \rho _{p} (t)|
n \rangle \{ I \pm \rho _{p} (t), | n ^{\prime} \rangle \langle n ^{\prime} |\},
\]  
when Eq. (A15) is considered together with Eq. (A14). The last two terms in the above 
equation just correspond to the relaxation term derived in section IV.
 
\section*{Appendix B}

Consider a partially filled band in a crystal. Because holes are vacencies
of any orbitals in this paper, the reference energy for holes is different
from that for particles. In a specific orbital, different wavefunctions are
used for particles and holes to obtain the correct crystal momentum, charge,
and the transition energy [10]. We diagonalize $\rho _{p}(t)$ as in Eq. (5), the
number of holes in $\lambda $ is $1-c_{\lambda }(t)$, so the density matrix
of holes is
\[
\rho _{h}(t)=\sum_{\lambda }[1-c_{\lambda }(t)]|\psi _{\lambda
}^{c}(t)\rangle \langle \psi _{\lambda }^{c}(t)|,
\ \ \ \ \ \ \ \ \ \ \ \ \ \ \ \ \ \ \ (B1)
\]
where $\psi _{\lambda }^{c}$ is the wavefunction for holes in orbital $%
\lambda $. It is easy to see that we can obtain $\rho _{h}(t)$ from $\rho _{%
\overline{p}}(t)$ after transforming the wavefunctions.

\section*{Appendix C}

To see that we cannot describe the gain of particles in orbital $|\phi
\rangle $ by Eq. (4), we just need to consider the special case when no
particle occupies $|\phi \rangle $ at $t=0$, i.e. $\langle \phi |\rho
_{p}(0)|\phi \rangle =0$. For any $|{\phi }^{\prime }\rangle $ orthogonal to
$|\phi \rangle $, we have
\[
|\langle \phi |\rho _{p}(0)|{\phi } ^{\prime} \rangle | \leq \langle \phi
|\rho _{p}(0)|\phi {\rangle }^{1/2}\langle {\phi }^{\prime }|\rho _{p}(0)|{%
\phi }^{\prime }{\rangle }^{1/2}=0,
\ \ \ \ \ \ \ \ \ \ \ \ \ \ \ \ \ \ \ (C1)
\]
from Cauchy's inequality, because $\rho _{p}(0)$ must be positive. Expanding $%
H(0)$, $A(0)$, and $\rho _{p}(0)$ in Eq. (4) by using any orthonormal
completet set containing $|\phi \rangle $, it is easy to see that $\frac{%
\partial }{\partial t}\langle \phi |\rho _{p}(0)|\phi \rangle =0$ for any $%
H(0)$ and $A(0)$. However, the gain rate should be the largest at $t=0$.

The change of the number of particles due to the loss and gain factors
should be proportional to $t$. If we take those terms of the order $%
t^{2} $, of course, the Liouville flow described by $H(t)$ can also induce
the increase of particles in $|\phi \rangle $.

\section*{Appendix D}

To see that the positivity of $\rho _{p}(t)$ can be broken by including the
pure-dephasing terms, consider a system in which only three orbitals $%
|1{\rangle }$, $|2{\rangle }$, and $|3{\rangle }$ are allowed, and assume that
the Hamiltonian $H_{0}$ is diagonalized by these three orbitals. Let the
equation for the time evolution of $\rho _{p}(t)$ be
\[
\frac{\partial }{\partial t}{\rho }_{p}(t)= \frac{1}{i} [ H _{0}, \rho_{p} (t)]- \Gamma
|2 \rangle \langle 2|{\rho }_{p}(t)|3\rangle \langle 3|-\Gamma |3\rangle \langle
3| \rho _{p}(t)|2\rangle \langle 2|. \ \ \ \ \ (D1)
\]
Assume, for example, that initially,
\[
{\rho }_{p}(0)=\frac{1}{3}(|1\rangle \langle 1|+|2\rangle \langle 2|+|3\rangle
\langle 3|) \ \ \ \ \ \ \ \ \ \ \ \ \ \ \ \ \ \ \ (D2)
\]
\[
+ \frac{10}{27} (|1 \rangle \langle 2| +|2\rangle \langle 1|+ |1\rangle \langle
3|+|3\rangle \langle 1| ) + \frac{2}{9}(|2\rangle \langle 3| + | 3 \rangle
\langle 2|).
\]
After calculations, it is easy to see that $\rho _{p}(0)$ is positive, but
one of the eigenvalue of $\rho _{p}$ becomes negative as $t\rightarrow
\infty $.

\end{document}